 \documentclass[11pt]{article}
 \usepackage[top=1in,bottom=1in,left=1.5in,right=1.5in]{geometry}
 \usepackage{amsmath, amssymb, amsfonts, enumerate}
  \newcommand{\hs}{\hspace*{\parindent}}

\begin{document}

 \title{For the Monomer-Dimer $\lambda_d(p)$, the Master Algebraic Conjecture}

 \author
 {Paul Federbush \\
 Department of Mathematics\\
 University of Michigan \\
 Ann Arbor, MI 48109-1043 \\
 (\texttt{pfed@umich.edu})}
 
  \date{\today}
 \maketitle

 \begin{abstract}
The author has recently presented two different expressions for $\lambda_d(p)$ of the monomer-dimer problem involving a power series in $p$, the first jointly with Shmuel Friedland. These two expressions are certainly equal, but this has not yet been proven rigorously. The first is naturally developed from quantities $\bar{J}_i$, cluster expansion kernels.  The second from the Mayer (or Virial) series of a dimer gas, in particular from the $b_i$ coefficients in the Mayer series. The sets $\{b_i\}$ and $\{\bar{J}_i\} $ can be derived from each other. Given an arbitrary set of values for either the $b_i$ or the $\bar{J}_i$ both expressions may be given in terms of a formal sum. The master algebraic conjecture is that these two expressions are equivalent. This is detailed in the special case all $\bar{J}_i$ are zero. 
 \end{abstract}

\section{ \underline{Introduction}}

Shmuel Friedland and the author recently presented an expression for  $\lambda_d (p)$  of the monomer-dimer problem involving a power series in $p$, see eq (7.1) of \cite{R1}. I have proven the series converges for small enough $p$, \cite{R2}. More recently I found another similar expression for  $\lambda_d (p)$ related in a suitable sense to the Mayer ( or Virial ) expansion of the dimer gas, \cite{R3}. It is known that this series converges for small enough $p$, \cite{R4}. The two series are certainly equivalent, but this has not yet been proven. It should be straight forward to prove the second expression converges to the correct value. The then two equivalent and seemingly very difficult tasks are either proving the two expressions for  $\lambda_d (p)$ are equivalent, or proving the first series converges to the correct value. It may first strike one that one should only work with the second expression, as being easier to construct proofs for, but it is the first expression with its development which incorporates the details of dependence on the dimension. We herein generalize the problem of the equivalence between the two expressions. We specify arbitrary values for the set of $\bar{J}_i$ ( or equivalently  the set of $b_i$ ) and then construct each of the two expressions ... for $\lambda_d(p)$ if the values were correctly picked ... now each involving formal power series. ( Whenever there is question all expressions in the paper are taken as formal expansions about zero in all variables. ) The \underline{Master Conjecture} is that the two expressions are equivalent, the formal sums are equivalent. To the author this result has some of the magical beauty that devotees of string theory sometimes see in their cult teachings. Section 2 displays the transition from the $\{b_i\}$ to the$\{\bar{J}_i\}$ and visa-versa. Section 3 constructs the first
expansion, and Section 4 the second expansion. Section 5 discusses the case when all $\bar{J}_i$ are zero. In Section 6 we present a  category theory formulation of the current master conjecture, we hope this is genuinely helpful.

\section{\underline{From $\{b_i\}$ to$\{\bar{J}_i\}$ and Visa-Versa}} 

We present the formulae we have used to derive the $\bar{J}_i$ from the $b_i$ in all our previous work, though they were not explicitly written down. They were implicitly mentioned in \cite{R5}.( See the remark there in the paragraph following the paragraph containing eq. (30), the sentence beginning: ``By a simple device...".) They follow upon consideration of the relation between the cluster expansion kernels developed with interaction ${\nu}$ and that with interaction ${f}$ as both considered in \cite{R5}.
We will \underline{always} impose the conditions

\begin{equation}
b_1=d
\end{equation}

\noindent
and

\begin{equation}
\bar{J}_1=0.
\end{equation}

\noindent
We first find $\bar{{J}^L}_r$, with $\bar{{J}^L}_1 = 0$,  and from $r=2$ on, inductively defined by

\begin{equation}
\bar{{J}^L}_r=\frac{1}{L}\left\{S_r-(\exp(L\sum_{i=1}^{r-1}\bar{{J}^L}_ix^i))|_r\right\}.
\end{equation}

\noindent
where

\begin{equation}
S_r=\sum_{p=0}^{r}\left\{(exp(L\sum_{i}b_i(\frac{x}{2d})^i))|_p \frac{1}{(r-p)!}\left(\frac{-1}{2(L-1)}\right)^{r-p}\frac{(L-2p)!}{(L-2r)!}\right\}.
\end{equation}

\noindent
The symbol $|$ with the subscript $j$ indicates the $j$th coefficient in the formal power series in x. The $\bar{J}_r$ are determined from the  $\bar{{J}^L}_r$ by taking $L$ to infinity. We may also inductively go from the $\bar{J}_i$ to the $b_i$ by the same formulae. We present now the explicit formulae thus obtained for the
first six terms going in each direction. We will several times give the first six terms explicitly for quantities to help the reader see relevant patterns, to suggest things they might want to prove.

\begin{equation}
\bar{J}_2=-\frac{1}{4}\frac{1}{d^2}(-2 d^2-b_2)
\end{equation}

\begin{equation}
\bar{J}_3=\frac{1}{24}\frac{1}{d^3}(3b_3+24 d b_2+28d^3)
\end{equation}

\begin{equation}
\bar{J}_4=-\frac{1}{16}\frac{1}{d^4}(-12d b_3-28 d^4-b_4-40 d^2 b_2)
\end{equation}

\begin{equation}
\bar{J}_5=\frac{1}{160}\frac{1}{d^5}(5 b_5+480 d^3 b_2+48 d^5-80 d b_2^2+420 d^2 b_3+80 d b_4)
\end{equation}

\begin{equation*}
\bar{J}_6=-\frac{1}{192}\frac{1}{d^6}(-1056 d^3 b_3+1584 d^4 b_2-432 d^2 b_4+1760 d^6-3 b_6 
\end{equation*}
\begin{equation}
+144 d b_3 b_2+1056 d^2 b_2^2-60 d b_5)
\end{equation}

\begin{equation}
b_2=d^2 (4\bar{J}_2-2)
\end{equation}

\begin{equation}
b_3= \frac{4}{3} d^3 (5 - 24 \bar{J}_2 + 6 \bar{J}_3 )
\end{equation}

\begin{equation}
b_4=2 d^4 (-14 +8 \bar{J}_4 +112 \bar{J}_2 - 48 \bar{J}_3 )
\end{equation}

\begin{equation}
b_5=\frac{16}{5} d^5 (42 +10 \bar{J}_5 -480\bar{J}_2 - 80 \bar{J}_4 + 270\bar{J}_3+80\bar{J}_2^2 )
\end{equation}

\begin{equation}
b_6=\frac{16}{3} d^6 (-132 -120 \bar{J}_5 +1980\bar{J}_2 +528 \bar{J}_4 -1320\bar{J}_3-1056\bar{J}_2^2 +12\bar{J}_6+288\bar{J}_3 \bar{J}_2)
\end{equation}

\section{\underline{The First Expression for $\lambda_d(p)$}}

Shmuel Friedland and the author presented the `first expression' for $\lambda_d (p)$ in \cite{R1}. We here write it
in the form

\begin{equation}
\lambda_d(p) = Q_1 + Q_2
\end{equation}

\begin{equation}
 Q_1 = \frac{1}{2} \left( p \ln (2d) - p \ln p - 2 (1-p) \ln (1-p) - p \right)
\end{equation}

\begin{equation}
Q_2 =  \sum_{k=2} a_k p^k.
\end{equation}

\noindent
We take our construction of the $a_i$ from \cite{R2}. First as in earlier work we define the $\alpha_i$. One solves the
following equation by iteration from $\alpha_i = 0$.

\begin{align}
\alpha_k = \left( \bar J_k p^k \right) \cdot \frac{1}{\left( 1 - 2 \sum i \alpha_i \right)^{2k}} \cdot \left( 1 - 2 \sum i \alpha_i / p \right)^{k}
\end{align}

\noindent
Then $Q_2$ is given as

\begin{align}
Q_2 =  \sum \alpha_i - \sum_{k=2} \frac{1}{k} \left( 2 \sum_i i \alpha_i \right)^k + \frac{1}{2} p \sum_{k=2} \frac{1}{k} \left( 2 \sum_i i \alpha_i / p \right)^k.
\end{align}

\noindent
We now list the values of the first few $a_i$, as derived from equations (17)  - (19).

\begin{equation}
a_2=\bar{J}_2
\end{equation}

\begin{equation}
a_3=-4\bar{J}_2^2+\bar{J}_3
\end{equation}

\begin{equation}
a_4=\bar{J}_4-12\bar{J}_2\bar{J}_3+\frac{80}{3}\bar{J}_2^3+8\bar{J}_2^2
\end{equation}

\begin{equation}
a_5=-128\bar{J}_2^3+\bar{J}_5+24\bar{J}_2\bar{J}_3-16\bar{J}_2\bar{J}_4-9\bar{J}_3^2-224\bar{J}_2^4+144\bar{J}_2^2\bar{J}_3
\end{equation}

\begin{equation*}
a_6=\frac{416}{3}\bar{J}_2^3+32\bar{J}_2\bar{J}_4+\frac{10752}{5}\bar{J}_2^5+18\bar{J}_3^2
\end{equation*}
\begin{equation*}
+1792\bar{J}_2^4+224\bar{J}_4\bar{J}_2^2+252\bar{J}_3^2\bar{J}_2
\end{equation*}
\begin{equation}
-1792\bar{J}_2^3\bar{J}_3+\bar{J}_6-672\bar{J}_2^2\bar{J}_3-24\bar{J}_3\bar{J}_4-20\bar{J}_2\bar{J}_5
\end{equation}

\section{\underline{The Second Expression for $\lambda_d(p)$}}

We follow the development of the `second expression' in \cite{R3}, in which this relation for $\lambda_d (p)$ was first
introduced. ( We there modestly cleped this expression `the Federbush relation'. ) One started with the Mayer series.( We will set $\beta = 1$ ).

\begin{equation} 
\beta P=\sum_{n=1}^\infty b_n z^n 
\end{equation}

\noindent
and the relation for p thought of as twice the density of dimers

\begin{equation}
p=2\sum_1^\infty n b_n z^n. 
\end{equation}
\noindent
or 

\begin{equation*}
z=\frac{1}{2b_1}p-\sum_2^\infty \frac{nb_n}{b_1}z^n.
\end{equation*}

\noindent
We solved this for $z = z(p)$ iterating from $z = 0$. Substituting this into the right side of eq.(25), one gets

\begin{equation}
P(p) = \sum_1^\infty b_n(z(p))^n
\end{equation}

\noindent
Written as a power series in $p$ this is the Virial series. We write $z(p)$ as

\begin{equation}
z=\frac{p}{2b_1}(1+F(p)).
\end{equation}

\noindent
Using the same notation as in eq. (15) and (16), the `second expression' has for $Q_2$ the relation

\begin{equation}
Q_2 = \frac{1}{2} \left(  2 (1-p) \ln (1-p) + p \right) +P(p) -\frac{p}{2} \ln(1 + F(p)).
\end{equation}

\noindent

We now list the expressions derived for the first few $a_i$ explicitly

\begin{equation}
a_2=\frac{1}{4} \frac{1}{d^2} (2 d^2+b_2)
\end{equation}

\begin{equation}
a_3=-\frac{1}{24} \frac{1}{d^4} (-4 d^4-3 b_3 d+6 b_2^2)
\end{equation}

\begin{equation}
a_4=\frac{1}{48} \frac{1}{d^6} (4 d^6-18 b_3 b_2 d+20 b_2^3+3 b_4 d^2)
\end{equation}

\begin{equation}
a_5=-\frac{1}{320} \frac{1}{d^8} (-10 d^3 b_5+45 d^2 b_3^2-360 b_2^2 d b_3+80 d^2 b_4 b_2+280 b_2^4-16 d^8)
\end{equation}

\begin{equation*}
a_6=\frac{1}{960} \frac{1}{d^{10}} (2016 b_2^5-180 b_3 d^3 b_4+840 d^2 b_4 b_2^2+32 d^{10}
\end{equation*}
\begin{equation}
+ 945 b_3^2 d^2 b2+15 b_6 d^4-150 b_5 b_2 d^3-3360 b_3 d b_2^3)
\end{equation}

\section{\underline{Special case, all $\bar{J}_i = 0$ }}

We first define $\tilde{b}_n$

\begin{equation}
\tilde{b}_n= \frac{1}{n} (-1) (-1)^n d^n 2^{n-1} C_n
\end{equation}

\noindent
where $C_n$ is the Catalan number

\begin{equation}
 C_n = \frac{(2n)!}{(n+1)! n!}
\end{equation}

\vspace{12pt}
\underline {Conjectured Theorem}, The set of $\bar{J}_i$ with all $\bar{J}_i=0$ corresponds to the set of $b_i$ with all $b_i = \tilde{b}_i$ under the transformations in Section 2. Both of the expressions for $\lambda_d(p)$ yield $Q_2 = 0$ in this case. 

\vspace{12pt}

\noindent
We break the proof into three parts, but we can prove only the first two parts.

\vspace{12pt}

\noindent
1) If all $\bar{J}_i=0$ expression one gives $Q_2 = 0$.

\noindent
2) If all $b_i = \tilde{b}_i$ then expression two gives $Q_2 = 0$.

\noindent
3) The set of all $\bar{J}_i=0$ corresponds to the set of all $b_i = \tilde{b}_i$.
 
 \vspace{12pt}
\noindent
We note that the truth of  1) is a trivial observation.

\vspace{12pt}
\noindent
We turn to the proof of 2), basically a number of computations. We write down the well known generating function for the Catalan numbers

\begin{equation}
 \sum_{n=0}^{\infty}C_n x^n = \frac{1-(1-4x)^{1/2}}{2x} = \frac{2}{1+(1-4x)^{1/2}}
\end{equation}

\noindent
If we substitute $\tilde{b}_n$ for $b_n$ into eq. (26) we get

\begin{equation}
 p = 1 - \sum_{0}^{\infty} (-2dz)^n C_n
\end{equation}

\noindent
We set $x = -2dz$ and then eq. (38) becomes

\begin{equation}
 1 - p =  \frac{1-(1-4x)^{1/2}}{2x} 
\end{equation}

\noindent
or

\begin{equation}
 x = - \frac{p}{(1-p)^2}
\end{equation}

\noindent
which becomes

\begin{equation}
 z = \frac{p}{2d(1-p)^2}
\end{equation}

\noindent
We substitute this into eq. (27) to get

\begin{equation}
 P(p) = -\frac{1}{2} \sum_{1}^{\infty} \frac{1}{n} \left( \frac{-p}{(1-p)^2}\right)^n C_n
\end{equation}

\noindent
A little calculation using the generating function leads to

\begin{equation}
 P = -\frac{p}{2} - \ln(1-p)
\end{equation}

\noindent
From eq. (28) and eq. (41) we directly get

\begin{equation}
1 + F = \frac{1}{(1-p)^2}
\end{equation}

\noindent
We now substitute eq. (43) and eq. (44) into eq. (29) and we get the desired result

\begin{equation}
Q_2 = 0.
\end{equation}

\vspace{12pt}
\noindent
The final task is to prove 3) above. We do not now know how to do this, but we write down what must be shown. We first find the expression for the sum in the exponential of eq. (4)

\begin{equation}
e(x) \equiv \sum_{i} \tilde{b}_i (\frac{x}{2d})^i = \frac{1}{1+(1+4x)^{1/2}} + \ln (1+(1+4x)^{1/2}) - \frac{1}{2} - \ln(2)
\end{equation}

\noindent
by a similar computation to that for eq. (43). Now what one has to prove is

\begin{equation}
0 =\lim_{L\to\infty}\frac{1}{L}\left\{\sum_{p=0}^{r}(exp(L e(x)))|_p \frac{1}{(r-p)!}\left(\frac{-1}{2(L-1)}\right)^{r-p}\frac{(L-2p)!}{(L-2r)!}\right\}
\end{equation}

\noindent
where the bar with subscript has the same meaning as in eq. (3) and eq. (4).

\vspace{12pt}
\noindent
Hopefully someone perhaps a little smarter than me, and certainly a lot younger, will prove this...and even the full master algebraic  conjecture.

\section{\underline{A Category Theory Formulation of the Problem}}

We work in the category whose elements are sequences of real numbers. An element $\bar{c}$ corresponds to a sequence

\begin{equation}
\bar{c} \leftrightarrow [c_1,c_2,..]
\end{equation}

\noindent
and we write $(\bar{c})_i = c_i$. The spaces we work with are each collections of such elements.  Each mapping $\phi$ will be required to have the property that for each i one has that
$\phi(\bar{c})_i$ is a polynomial in $\{c_1,c_2,...,c_i\}$. We note that a composition of such mappings also has this property, something that must be true for consistency with the key property of categories.

\vspace{12pt}
\noindent
We consider as elements of our spaces the sequences we now call $\bar{j}$, $\bar{b}$, $\bar{a}$ and $\bar{a'}$ which are as follows

\begin{equation}
\bar{j} \leftrightarrow[\bar{J}_1,\bar{J}_2...]
\end{equation}

\begin{equation}
\bar{b} \leftrightarrow[b_1,b_2...]
\end{equation}

\begin{equation}
\bar{a} \leftrightarrow [a_1,a_2...]
\end{equation}

\begin{equation}
\bar{a'} \leftrightarrow [a'_1,a'_2...]
\end{equation}

\vspace{12pt}
\noindent
We let $f$ be the mapping carrying $\bar{b}$ to $\bar{j}$ 

\begin{equation}
f: \bar{b} \rightarrow \bar{j}
\end{equation}

\noindent
defined by eqs.(3) and (4) and whose first six components are detailed in eqs. (2), (5)-(9).

\vspace{12pt}
\noindent
We let $f ^{-1}$ be the mapping carrying $\bar{j}$ to $\bar{b}$ 

\begin{equation}
f^{-1}: \bar{j} \rightarrow \bar{b}
\end{equation}

\noindent
defined by eqs.(3) and (4) and whose first six components are detailed in eqs. (1), (10)-(14).

\vspace{12pt}
\noindent
We let g be the mapping carrying $\bar{j}$ to $\bar{a}$ 

\begin{equation}
g: \bar{j} \rightarrow \bar{a}
\end{equation}

\noindent
defined by eqs.(17) - (19) and whose first six components are detailed in eqs. (20) - (24), and the definition $a_1 = 0$.

\vspace{12pt}
\noindent
We let h be the mapping carrying $\bar{b}$ to $\bar{a'}$ 

\begin{equation}
h: \bar{b} \rightarrow \bar{a'}
\end{equation}

\noindent
defined by eqs.(26) - (29) and whose first six components are detailed in eqs. (30) - (34) ( we now prime the $a_i$'s of these equations to distinguish them from those of the previous mapping ), and the definition $a'_1 = 0$. ( $g$ and $h$ in fact have inverses. )

\vspace{12pt}
\noindent
The master algebraic conjecture is then simply the statement

\begin{equation}
h \circ f ^{-1}= g
\end{equation}

\noindent
or

\begin{equation}
h  = g \circ f.
\end{equation}

\end{document}